\pdfoutput=1
\documentclass[11pt]{article}

\usepackage[]{ACL2023}

\usepackage{times}
\usepackage{latexsym}
\usepackage{booktabs}
\usepackage{comment}
\usepackage{graphicx}
\usepackage{amsmath}
\usepackage[T1]{fontenc}
\usepackage[utf8]{inputenc}

\usepackage{microtype}

\usepackage{inconsolata}

\title{A2TTS: TTS for Low Resource Indian Languages}

\author{
  Ayush Singh Bhadoriya\quad
  Abhishek Nikunj Shinde\quad
  Isha Pandey\quad \\
  \textbf{Ganesh Ramakrishnan} \\
  Indian Institute of Technology Bombay \\
}
  
\begin{document}
\maketitle
\begin{abstract}
    We present a speaker conditioned text-to-speech (TTS) system aimed at addressing challenges in generating speech for unseen speakers and supporting diverse Indian languages. Our method leverages a diffusion-based TTS architecture, where a speaker encoder extracts embeddings from short reference audio samples to condition the DDPM decoder for multi-speaker generation. To further enhance prosody and naturalness, we employ a cross-attention-based duration prediction mechanism that utilizes reference audio, enabling more accurate and speaker-consistent timing. This results in speech that closely resembles the target speaker while improving duration modeling and overall expressiveness. Additionally, to improve zero-shot generation, we employed classifier-free guidance, allowing the system to generate speech more near speech for unknown speakers. % We present a speaker-condition text-to-speech (TTS) system designed to address challenges in generating speech for unknown speakers and diverse Indian languages, where current systems often fall short. We utilize a diffusion-based TTS system, and incorporate a speaker encoder to extract speaker embeddings from a small sample of target speaker audio, which is used to condition the DDPM decoder for multi-speaker TTS. We effectively captured acoustic features for speakers in the training data but initially struggled with zero-shot generation for unseen speakers and obtaining speaker dependent durations for given text. To address both of the issues, we used an attention-based mechanism that extracts durations by leveraging a 2-second mel spectrogram from a different audio sample of the same speaker and text embeddings, as the attention mechanism takes both as input. This helps improve prosody by extracting prosodic features from the reference audio, leading to more natural speech with better duration predictions. This improves the accuracy of duration modeling, allowing for more natural speech timing because, the predicted durations are influenced by speaker characteristics, enabling better adaptation to different voices. Additionally, to improve zero-shot generation, we employed classifier-free guidance, allowing the system to generate speech more near speech for unknown speakers.
    Using this approach, we trained language-specific speaker-conditioned models using the IndicSUPERB dataset for multiple Indian languages such as Bengali, Gujarati, Hindi, Marathi, Malayalam, Punjabi and Tamil.
\end{abstract}

\begin{figure*}[h]
    \centering
    % \includesvg[width=0.9\textwidth]{tts_architecture.drawio_1.svg}
    \includegraphics[width=\linewidth]{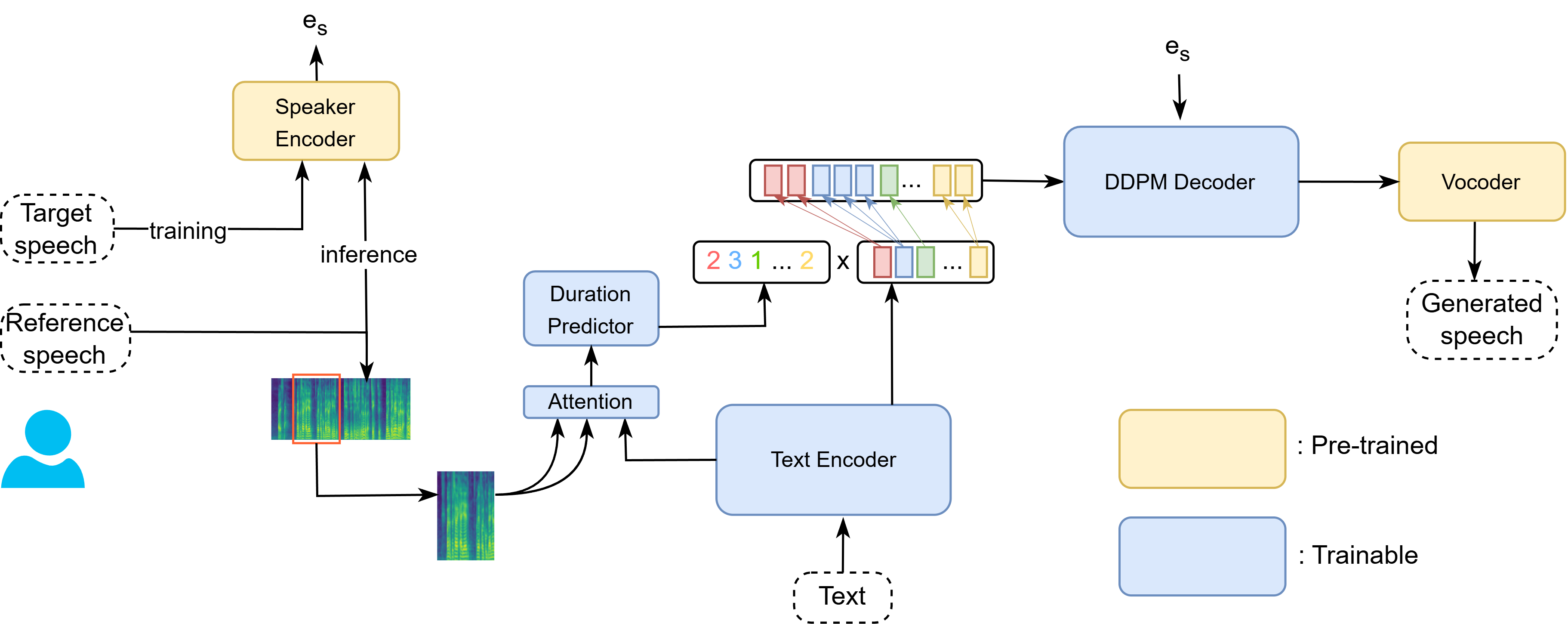}
    \caption{Overview of TTS architecture. During training, the model utilizes two different audio samples from the same speaker to enhance prosody learning and speaker adaptation. The target speech contains the target transcript used to generate speaker embedding \( e_s \), while the reference audio does not contain the same text but provides additional speaker-specific information for duration prediction. During inference, only a single reference audio is used to generate the speaker embedding and extract prosody from its mel spectrogram}
    \label{fig:architecture}
\end{figure*}

\section{Introduction}

% Text-to-speech (TTS) systems have evolved rapidly, moving from rule-based synthesis to end-to-end neural models capable of generating highly natural speech. Foundational architectures such as Tacotron 2 \cite{shen2018natural}, FastSpeech \cite{ren2019fastspeech}, and Glow-TTS \cite{kim2020glow} introduced attention-based and flow-based techniques for efficient speech generation. More recently, diffusion-based models like Grad-TTS \cite{popov2021grad} have shown superior performance in producing high-quality, prosodically rich speech by modeling the generation process as iterative denoising.
Text-to-speech (TTS) systems have evolved rapidly, moving from rule-based synthesis to end-to-end neural models capable of generating highly natural speech. Foundational architectures such as Tacotron 2~\cite{shen2018natural}, FastSpeech~\cite{ren2019fastspeech}, and Glow-TTS~\cite{kim2020glow} introduced attention-based and flow-based techniques for efficient speech generation. Subsequent models such as FastSpeech 2~\cite{ren2020fastspeech}, VITS~\cite{kim2021conditional}, and Grad-TTS~\cite{popov2021grad} further improved quality and inference speed through non-autoregressive and variational approaches. More recently, diffusion-based models like FastDiff~\cite{huang2022fastdiff} and ProDiff~\cite{huang2023prodiff} have shown strong performance in naturalness and robustness, while large-scale systems such as Voicebox~\cite{le2023voicebox} and F5-TTS~\cite{chen2023f5} push toward few-shot and multilingual, zero-shot synthesis capabilities.

Despite these advancements, achieving zero-shot speaker adaptation synthesizing speech for unseen speakers without finetuning remains a key challenge. Approaches such as YourTTS \cite{casanova2022yourtts} and methods using x-vectors \cite{snyder2018x} or d-vectors have attempted to generalize across speakers using learned embeddings. UnitSpeech \cite{kim2023unitspeech} built upon Grad-TTS by incorporating a speaker encoder to enable speaker-conditioned diffusion, improving zero-shot performance without requiring explicit speaker labels.

In this work, we extend the Grad-TTS framework to further improve speaker adaptation and prosody modeling. We adopt the speaker encoder and conditioning mechanism from UnitSpeech to extract embeddings from short reference audio. These embeddings condition both the duration predictor and the diffusion decoder, enabling high-quality, zero-shot voice synthesis. 
% Our key modification is the integration of a cross-attention mechanism in the duration predictor, allowing it to incorporate prosodic cues from a fixed-length reference mel spectrogram. This helps capture speaker-specific timing, resulting in more expressive and rhythmically accurate speech.
Our key modification is the integration of an end-to-end cross-attention-based duration predictor, which eliminates the need for explicitly training a separate duration model, as required in systems like Voicebox\cite{le2023voicebox} and F5-TTS\cite{chen2023f5}.

During inference, we employ classifier-free guidance (CFG) \cite{ho2022classifier}, following the implementation from UnitSpeech \cite{kim2023unitspeech}, where the unconditional condition is represented by the dataset-wide mel-spectrogram mean. This enhances pronunciation and speaker control without modifying the training procedure.

% We evaluated our models in the IndicSUPERB data set  \cite{https://doi.org/10.48550/arxiv.2208.11761} , which includes nine Indian languages and Vistaar benchmark dataset for wers. Our approach demonstrates strong speaker generalization and prosody preservation across diverse speakers and low-resource language scenarios.
We evaluated our models on the IndicSUPERB dataset~\cite{https://doi.org/10.48550/arxiv.2208.11761}, which includes nine Indian languages, and on the Vistaar benchmark dataset\cite{ganesh2023vistaar} for WER evaluation. We chose to train our models exclusively on the IndicSUPERB dataset due to its high-quality, noise-free recordings and its diversity in speakers, accents, and speaking styles. These characteristics make it particularly well-suited for building robust, speaker-conditioned TTS systems, especially in low-resource language settings. Our approach demonstrates strong speaker generalization and prosody preservation across diverse speakers and linguistic scenarios.

\section{Methodology}

% To achieve high-quality audio generation while preserving speaker characteristics and prosody, we propose an architecture that systematically processes input text through a series of neural network modules, leveraging attention mechanisms and diffusion models. The approach ensures accurate duration prediction and realistic mel spectrogram synthesis, conditioned on speaker embeddings. Our methodology consists of three key components: text encoding, duration prediction, and diffusion-based mel spectrogram refinement.
We build upon Grad-TTS, the diffusion-based text-to-speech architecture also used in UnitSpeech, as the foundational model for our system. To preserve speaker identity and prosody, we adopt the speaker conditioning approach introduced in UniSpeech, enabling effective integration of speaker embeddings throughout the generation process. While retaining the original components of Grad TTS namely text encoding, duration prediction, and diffusion-based mel spectrogram synthesis our primary modification lies in the duration prediction module, where we introduce a cross-attention mechanism. This enhancement allows the duration predictor to better align linguistic features with speaker characteristics, leading to improved temporal modeling and more natural speech synthesis.

\subsection{Text Encoding}

% The first stage in our architecture involves converting input text into text embeddings that effectively represent phonemes and their relationships. The input text is passed through a text encoder, which transforms it into a sequence of dense vector representations:
We retain the text encoding mechanism from the original Grad-TTS architecture, as it effectively captures the phonetic and linguistic structure necessary for high-quality speech synthesis. In this stage, the input text typically represented as a sequence of phonemes is processed by a text encoder to generate dense vector representations that encapsulate contextual relationships among phonemes.

\begin{equation}
E_t = f_{\text{encoder}}(T),
\end{equation}

where \(T\) denotes the input phoneme sequence, and \(E_t\) represents the resulting text embeddings. These embeddings form the core linguistic features used in downstream modules, including duration prediction and mel spectrogram generation.

\subsection{Duration Prediction}

Accurate duration prediction is essential for producing speech with natural rhythm and speaker-specific prosody. In our architecture, we build upon the original duration predictor in Grad-TTS by introducing a \textit{cross-attention mechanism} between the encoded text and a reference mel spectrogram. This enhancement allows the model to align phoneme durations more closely with speaker-dependent prosodic patterns.

The text embeddings \(E_t\), generated by the text encoder, are used as queries in a cross-attention module, while a fixed-length reference mel spectrogram \(M\) from the same speaker representing a 2-second audio segment unrelated to the input text is used as keys and values:

\begin{equation}
    A = \text{Attention}(E_t, M, M),
\end{equation}

where:
- \( A \) is the attention \cite{vaswaniattention} output,
- \( M \) is the reference mel spectrogram,
- \( E_t \) is used as queries, and \( M \) serves as both keys and values.

To avoid overfitting, the reference mel spectrogram is chosen to be unrelated to the input text. If the reference and input share the same linguistic content, the model tends to memorize durations and produce incorrect alignments. By using unrelated content, the model instead focuses on capturing speaker-specific timing and intonation patterns.

The attention output \(A\) is then passed to the duration prediction network:

\begin{equation}
    D = f_{\text{dur}}(A),
\end{equation}

where \(D\) represents the predicted duration for each phoneme. This modification enables our model to produce speaker-adaptive durations, contributing to more expressive and natural-sounding speech synthesis.

\subsection{Diffusion-Based Mel-Spectrogram Refinement}

In the final stage of our model, we employ a denoising diffusion probabilistic model (DDPM) to generate high-quality mel spectrograms from an intermediate noisy representation. The decoder architecture follows the same design as Grad-TTS, where the model iteratively refines an initial spectrogram \(X_0\) through a reverse diffusion process parameterized by a learnable score function \(s_\theta(X_t, t)\).

To condition the model on speaker identity, we adopt the speaker conditioning strategy used in UnitSpeech. Specifically, we utilize UnitSpeech’s pre-trained speaker encoder to extract speaker embeddings \(e_s\), which are used to condition the score function as \(s_\theta(X_t, t \mid c_{\text{c}}, e_s)\), where \(c_{\text{c}}\) represents the aligned output of the text encoder.

During inference, we further enhance the expressiveness and speaker consistency of the generated speech using classifier-free guidance (CFG) \cite{ho2022classifier}, implemented as in UnitSpeech. Instead of learning an unconditional embedding, we use the dataset-wide mean mel spectrogram \(c_{\text{mel}}\) as the unconditional condition. The modified score function becomes:

\begin{equation}
    \hat{s}_\theta(X_t, t \mid c_{\text{c}}, e_s) = s_\theta(X_t, t \mid c_{\text{c}}, e_s) + \gamma \cdot \alpha_t,
\end{equation}
\begin{equation}
    \alpha_t = s_\theta(X_t, t \mid c_{\text{c}}, e_s) - s_\theta(X_t, t \mid c_{\text{mel}}, e_s),
\end{equation}

where \(\gamma\) is a guidance scale factor that controls the strength of conditioning. This inference-time modification amplifies the influence of the speaker and text conditions without requiring changes to the training process, leading to clearer pronunciation and improved speaker consistency, especially for unseen voices.

% \begin{table*}
% \centering
% \resizebox{0.8\textwidth}{!}{%
% \begin{tabular}{lccccccc}
% \hline
% \toprule
% \textbf{Dataset} & \textbf{Hindi} & \textbf{Punjabi} & \textbf{Marathi} & \textbf{Gujarati} & \textbf{Bengali} & \textbf{Malayalam} & \textbf{Tamil} \\
% \hline
% \midrule
% Indicsuperb   & 6.16\%  & 7.93\%  & 6.00\%  & 7.31\%  & 6.90\%  & 12.25\% & 10.10\%  \\
% Fleurs        & 8.37\%  & 9.59\%  & 12.76\% & 9.87\%  & 13.91\% & 16.72\% & 12.53\% \\
% CommonVoice   & 11.67\% & 12.61\% & 12.24\% & --      & 11.08\% & 17.93\% & 15.16\% \\
% IndicTTS      & 10.07\% & --      & 14.00\% & 12.36\% & 16.35\% & 7.19\%  & 6.33\% \\
% Kathbath      & 8.52\%  & 7.40\%  & 10.79\% & 8.75\%  & 12.90\% & 12.86\% & 10.19\% \\
% \bottomrule
% \hline
% \end{tabular}%

% }
% \caption{Character Error Rates (CER) across different datasets and languages. All language models were trained solely on the IndicSUPERB dataset, except for Malayalam and Tamil, which also included the IndicTTS dataset during training. This additional data resulted in lower CERs for Malayalam and Tamil on the IndicTTS evaluation set.}
% \label{tab:cer_table}
% \end{table*}

\begin{table*}
\centering
\resizebox{\textwidth}{!}{%
\begin{tabular}{lccccccccccccccc}
\toprule
\textbf{Dataset} & \multicolumn{2}{c}{\textbf{Hindi}} & \multicolumn{2}{c}{\textbf{Punjabi}} & \multicolumn{2}{c}{\textbf{Marathi}} & \multicolumn{2}{c}{\textbf{Gujarati}} & \multicolumn{2}{c}{\textbf{Bengali}} & \multicolumn{2}{c}{\textbf{Malayalam}} & \multicolumn{2}{c}{\textbf{Tamil}} \\
\cmidrule(lr){2-3} \cmidrule(lr){4-5} \cmidrule(lr){6-7} \cmidrule(lr){8-9} \cmidrule(lr){10-11} \cmidrule(lr){12-13} \cmidrule(lr){14-15}
& \textbf{TTS} & \textbf{ASR} 
 & \textbf{TTS} & \textbf{ASR} 
 & \textbf{TTS} & \textbf{ASR} 
 & \textbf{TTS} & \textbf{ASR} 
 & \textbf{TTS} & \textbf{ASR} 
 & \textbf{TTS} & \textbf{ASR} 
 & \textbf{TTS} & \textbf{ASR} \\
\midrule
Indicsuperb    & 6.16 & 2.50 & 7.93 & 3.46 & 6.00 & 2.77 & 7.31 & 2.88 & 6.90 & 1.79 & 12.25 & 4.92 & 10.10 & 3.51 \\
\cite{indicSUPERB2022}\\
Fleurs         & 8.37 & 4.81 & 9.59 & 5.11 & 12.76 & 4.73 & 9.87 & 4.74 & 13.91 & 4.34 & 16.72 & 5.22 & 12.53 & 11.95 \\
\cite{conneau2023fleurs}\\
CommonVoice    & 11.67 & 3.93 & 12.61 & 3.88 & 12.24 & 3.65 & --   & --   & 11.08 & 3.40 & 17.93 & 6.59 & 15.16 & 4.32 \\
\cite{ardila2020common}\\
IndicTTS       & 10.07 & 1.82 & --   & --   & 14.00 & 2.58 & 12.36 & 3.27 & 16.35 & 2.23 & 7.19 & 1.70 & 6.33 & 2.36 \\
\cite{indictts}\\
Kathbath       & 8.52 & 2.99 & 7.40 & 3.47 & 10.79 & 5.61 & 8.75 & 3.65 & 12.90 & 3.70 & 12.86 & 4.85 & 10.19 & 3.53 \\
\cite{kathbath2022}\\
\bottomrule
\end{tabular}%
}
\caption{
Character Error Rates (CER) across different datasets and languages. For each language, two subcolumns are shown: TTS Error is the CER produced by the ASR model on synthesized speech, and ASR error is the CER when decoding the original reference transcriptions (i.e., ground-truth synthesis CER). All models were trained solely on the IndicSUPERB dataset, except for Malayalam and Tamil, which also included IndicTTS\cite{indictts} during training, improving ASR performance on the IndicTTS evaluation set.
}
\label{tab:cer_table}
\end{table*}

\section{Experiments}
\subsection{Dataset}
We utilized the IndicSUPERB dataset to train our multi-speaker text-to-speech (TTS) model. This dataset encompasses 12 Indian languages, each contributing a substantial amount of labeled speech data. For language-specific TTS models, we focused on individual languages within the dataset.

The audio recordings in the IndicSUPERB dataset have an average duration of approximately 6 seconds per utterance. Originally sampled at 16 kHz, we resampled the audio data to 22 kHz to enhance the quality of our TTS models.

For testing, IndicSUPERB offers a dedicated dataset comprising recordings from 20 speakers per language (10 male and 10 female), totaling approximately 3 hours of audio per language. These speakers are distinct from those in the training set, ensuring an unbiased evaluation of our models.

Specifically, we employ the speaker encoder from UnitSpeech, which is trained on VoxCeleb2 \cite{chung2018voxceleb2} a dataset comprising 6,112 speakers. For the vocoder, we adopt the same approach as Grad-TTS, utilizing the HiFi-GAN \cite{kong2020hifi}, which is known for delivering high-quality audio synthesis. We apply this vocoder consistently across all models in our comparisons.

In preprocessing, for each speaker, we extracted audio samples to obtain speaker embeddings. Additionally, we selected a separate 2-second segment from the same speaker's audio to input into the attention mechanism for duration prediction. This approach prevents the attention mechanism from overfitting and producing inaccurate durations when the same audio is used for both embedding extraction and attention input.

\subsection{Training and Fine-tuning}
Initially, we trained the TTS model, which shares the same architecture as Grad-TTS but with double the number of channels to accommodate multi-speaker modeling, for 1,500 epochs. To enhance prosody, we incorporated additional information from a mel spectrogram of the same speaker but from a different audio sample than the text input. This involved applying attention mechanisms over the mel spectrogram and text embeddings generated by the text encoder. We fine-tuned this setup for an additional 1,000 epochs.

Training and fine-tuning were conducted on 8 NVIDIA H100 GPUs. We employed the Adam optimizer with a learning rate of $1 \text{e}^{-4}$ and a batch size of 64.

\section{Results}
\subsection{Evaluation Metrics}

We use two primary metrics to evaluate our TTS system:

Sim-O Score measures speaker similarity between reference and synthesized speech via cosine similarity of speaker embeddings. Higher scores (closer to 1) indicate better speaker identity preservation.

Character Error Rate (CER) evaluates intelligibility by comparing ASR transcriptions of synthesized speech to the original text. We report CER as the main intelligibility metric, since baseline ASR Conformer models yield high Word Error Rates (WER) even on ground-truth audio, making WER less reliable for assessing TTS performance.

To complement this, WER scores are provided in the Appendix for both original and synthesized speech across datasets and languages.

\begin{table}[h!]
\centering
% \begin{tabular}{lccc}
\begin{tabular}{l p{2.2cm} p{1cm} p{1.3cm}}
% \begin{tabular}{l p{2.5cm}<{\centering} p{1cm}<{\centering} p{1.5cm}<{\centering}}
% \begin{tabular}{lccc}
\hline
% \toprule
\textbf{Language} & \textbf{Gradtts \t  conditioned on speaker embeddings} & \textbf{A2TTS} & \textbf{Test set Average} \\
\hline
% \midrule
Hindi       & 0.7253  & 0.7291  & 0.7472  \\
Marathi     & 0.7157  & 0.7336  & 0.7714  \\
Gujarati    & 0.7073  & 0.7221  & 0.7250  \\
Punjabi     & 0.6905  & 0.7046  & 0.7717  \\
Bengali     & 0.6829  & 0.7082  & 0.7540  \\
Tamil       & 0.6748  & 0.7264  & 0.7534  \\
Malayalam   & 0.6921  & 0.7148  & 0.7721  \\
% Kannada     & 0.7029  & 0.7442  & 0.7605  \\
% Telugu      & 0.6822  & 0.6880  & 0.7701  \\
\hline
% \bottomrule
\end{tabular}

\caption{SIM-O Evaluation Scores for Different Languages.}
\label{tab:recognition_eval}
\end{table}

Here the gradtts conditioned on speaker embeddings is presented as baseline and test set average is the simo over the test dataset of indicsuperb.

\section{Conclusions}
In this work, we introduced a speaker-adaptive, diffusion-based text-to-speech (TTS) system capable of generating high-quality, natural speech for both seen and unseen speakers across multiple low-resource Indian languages. By incorporating a speaker encoder and classifier-free guidance, our model effectively captures speaker characteristics and improves zero-shot speaker adaptation. Additionally, our attention-based speaker-dependent duration modeling approach enhances prosody by leveraging a reference mel spectrogram to predict more natural speech timing, enabling better speaker adaptation.

Through extensive experiments using the IndicSUPERB dataset, we demonstrated that our approach significantly improves speaker similarity, prosody, and naturalness of synthesized speech across diverse Indian languages, including Bengali, Gujarati, Hindi, Marathi, Malayalam, Punjabi, and Tamil. The proposed methodology outperforms existing models in zero-shot speaker adaptation while maintaining high fidelity in speaker identity and expressiveness.

Our results indicate that leveraging diffusion models with guided speaker adaptation provides a scalable solution for speaker-personalized TTS systems. Future work can explore extending this approach to additional languages and further refining prosody modeling techniques to enhance expressiveness. Additionally, integrating more robust speaker adaptation techniques could further improve performance in real-world applications such as personalized virtual assistants, multilingual accessibility, and content creation.

\section*{Limitations}

While our TTS system demonstrates strong performance across multiple languages, it has several limitations:

Our models are trained only on IndicSUPERB; adapting to out-of-domain speakers or languages may require fine-tuning with additional data.

Limited Speaker Adaptability: The model is currently speaker-conditioned and not fine-tuned on specific speakers. However, with fine-tuning, it could potentially exhibit speaker adaptation capabilities, enabling closer voice matching for new target speakers.

Restricted Speaker Diversity: Training is limited to the IndicSUPERB dataset, which may not capture the full range of speaker variability. As a result, the model struggles to mimic speaker characteristics for voices that are significantly different from those seen during training.

High Training Cost: The model requires a large number of training epochs (approximately 2000) to converge effectively. This results in slow training times and demands resource, which may limit scalability and experimentation.

% In addition, limitations such as low scalability to long text, the requirement of large GPU resources, or other things that inspire crucial further investigation are welcome.

% \section*{Ethics Statement}
% Scientific work published at ACL 2023 must comply with the ACL Ethics Policy.\footnote{\url{https://www.aclweb.org/portal/content/acl-code-ethics}} We encourage all authors to include an explicit ethics statement on the broader impact of the work, or other ethical considerations after the conclusion but before the references. The ethics statement will not count toward the page limit (8 pages for long, 4 pages for short papers).

\section*{Acknowledgements}
We would like to thank Pranav Gaikwad for his invaluable guidance and support throughout this project. We are also grateful to the BharatGen team for providing the compute resources that enabled this research. Special thanks to the broader AI4Bharat initiative and the teams behind IndicSUPERB and IndicTTS, whose datasets and tools were instrumental to our experiments. Finally, we appreciate the contributions of all collaborators and reviewers who provided valuable insights and feedback during the course of this work.

\bibliography{custom}
\bibliographystyle{acl_natbib}

% \appendix
\appendix
\section*{Appendix}
\addcontentsline{toc}{section}{Appendix}
\section{WER Scores on ASR Conformer Outputs and TTS-Generated Speech}
The following tables show the Word Error Rates (WER) of ASR conformer models evaluated on the original test audio, as well as on the speech synthesized by our TTS system. These results are reported for multiple Indian languages from the Vistaar benchmark dataset. The WER scores are not included in the main paper due to the relatively high error rates of the baseline ASR systems.

% \vspace{1em}

% -------- Tables 1 and 2 (Hindi, Marathi) --------
\begin{table}[h!]
% \begin{table}[!htbp]
\centering
\small
\begin{minipage}[t]{0.35\linewidth}
\centering
\caption{WER (\%) for Hindi}
\begin{tabular}{lcc}
\toprule
\textbf{Dataset} & ASR & TTS \\
\midrule
IndicSuperb & 8.27 & 15.73 \\
CommonVoice & 10.99 & 26.12 \\
FLEURS & 11.39 & 18.08 \\
IndicTTS & 7.68 & 25.43 \\
MUCS & 9.28 & 25.97 \\
Kathbath & 8.85 & 18.64 \\
\bottomrule
\end{tabular}
\end{minipage}
\hfill
\begin{minipage}[t]{0.35\linewidth}
\centering
\caption{WER (\%) for Marathi}
\begin{tabular}{lcc}
\toprule
\textbf{Dataset} & ASR & TTS \\
\midrule
IndicSuperb & 12.34 & 21.61 \\
CommonVoice & 16.38 & 39.02 \\
FLEURS & 16.80 & 34.58 \\
IndicTTS & 12.01 & 41.62 \\
MUCS & 15.74 & 56.33 \\
Kathbath & 16.92 & 30.03 \\
\bottomrule
\end{tabular}
\end{minipage}
\end{table}

% -------- Tables 3 and 4 (Punjabi, Gujarati) --------
% \begin{table}[h!]
\begin{table}[!htbp]
\centering
\small
\begin{minipage}[t]{0.35\linewidth}
\centering
\caption{WER (\%) for Punjabi}
\begin{tabular}{lcc}
\toprule
\textbf{Dataset} & ASR & TTS \\
\midrule
IndicSuperb & 11.80 & 20.99 \\
CommonVoice & 13.35 & 32.72 \\
FLEURS & 14.93 & 24.75 \\
IndicTTS & -- & -- \\
MUCS & -- & -- \\
Kathbath & 11.83 & 20.46 \\
\bottomrule
\end{tabular}
\end{minipage}
\hfill
\begin{minipage}[t]{0.35\linewidth}
\centering
\caption{WER (\%) for Gujarati}
\begin{tabular}{lcc}
\toprule
\textbf{Dataset} & ASR & TTS \\
\midrule
IndicSuperb & 12.42 & 22.67 \\
CommonVoice & -- & -- \\
FLEURS & 16.49 & 27.24 \\
IndicTTS & 14.78 & 37.43 \\
MUCS & 21.98 & 26.79 \\
Kathbath & 13.55 & 24.95 \\
\bottomrule
\end{tabular}
\end{minipage}
\end{table}

% -------- Tables 5 and 6 (Bengali, Tamil) --------
% \begin{table}[h!]
\begin{table}[!htbp]
\centering
\small
\begin{minipage}[t]{0.35\linewidth}
\centering
\caption{WER (\%) for Bengali}
\begin{tabular}{lcc}
\toprule
\textbf{Dataset} & ASR & TTS \\
\midrule
IndicSuperb & 9.35 & 23.63 \\
CommonVoice & 14.03 & 35.13 \\
FLEURS & 15.45 & 34.91 \\
IndicTTS & 15.02 & 48.33 \\
MUCS & -- & -- \\
Kathbath & 12.24 & 33.21 \\
\bottomrule
\end{tabular}
\end{minipage}
\hfill
\begin{minipage}[t]{0.35\linewidth}
\centering
\caption{WER (\%) for Tamil}
\begin{tabular}{lcc}
\toprule
\textbf{Dataset} & ASR & TTS \\
\midrule
IndicSuperb & 20.69 & 38.43 \\
CommonVoice & 25.76 & 56.28 \\
FLEURS & 27.18 & 41.52 \\
IndicTTS & 16.84 & 29.33 \\
MUCS & 20.98 & 41.13 \\
Kathbath & 20.70 & 38.25 \\
\bottomrule
\end{tabular}
\end{minipage}
\end{table}

% -------- Table 7 (Malayalam) --------
% \begin{table}[h!]
\begin{table}[!htbp]
\centering
\small
\caption{WER (\%) for Malayalam}
\begin{tabular}{lcc}
\toprule
\textbf{Dataset} & ASR & TTS \\
\midrule
IndicSuperb & 29.22 & 47.88 \\
CommonVoice & 35.99 & 58.11 \\
FLEURS & 20.57 & 56.19 \\
IndicTTS & 17.76 & 36.97 \\
MUCS & -- & -- \\
Kathbath & 29.11 & 48.99 \\
\bottomrule
\end{tabular}
\end{table}

\end{document}